\documentclass[conference]{IEEEtran}
\usepackage{cite}
\usepackage{amsmath,amssymb,amsfonts}
\usepackage{algorithmic}
\usepackage{graphicx}
\usepackage{textcomp}
\usepackage{xcolor}

\usepackage{gensymb} 
\usepackage{multirow} 
\usepackage{booktabs} 
\usepackage{listings} 
\usepackage{framed} 
\usepackage[hidelinks]{hyperref}

\newcommand{\orcid}[1]{ORCID: \href{https://orcid.org/#1}{#1}}

\def\BibTeX{{\rm B\kern-.05em{\sc i\kern-.025em b}\kern-.08em
    T\kern-.1667em\lower.7ex\hbox{E}\kern-.125emX}}
\begin{document}

\title{Identifying Usability Issues of Software Analytics Applications in Immersive Augmented Reality}

\author{\IEEEauthorblockN{David Baum\IEEEauthorrefmark{1}, Stefan Bechert\IEEEauthorrefmark{2}, Ulrich Eisenecker\IEEEauthorrefmark{3}, Isabelle Meichsner\IEEEauthorrefmark{4}, and Richard M{\"u}ller\IEEEauthorrefmark{5}}
	\IEEEauthorblockA{ Information Systems Institute\\
		Leipzig University, Leipzig, Germany\\
	\IEEEauthorrefmark{1}Email: david.baum@uni-leipzig.de}
	\IEEEauthorblockA{\IEEEauthorrefmark{2}\orcid{0000-0002-6727-7106}}
	\IEEEauthorblockA{\IEEEauthorrefmark{3}Email: eisenecker@wifa.uni-leipzig.de}
	\IEEEauthorblockA{\IEEEauthorrefmark{4}\orcid{0000-0002-9474-382X}}
	\IEEEauthorblockA{\IEEEauthorrefmark{5}\orcid{0000-0001-6730-4082}}}

\maketitle

\begin{abstract}
Software analytics in augmented reality (AR) is said to have great potential.
One reason why this potential is not yet fully exploited may be usability problems of the AR user interfaces.
We present an iterative and qualitative usability evaluation with 15 subjects of a state-of-the-art application for software analytics in AR.
We could identify and resolve numerous usability issues.
Most of them were caused by applying conventional user interface elements, such as dialog windows, buttons, and scrollbars.
The used city visualization, however, did not cause any usability issues.
Therefore, we argue that future work should focus on making conventional user interface elements in AR obsolete by integrating their functionality into the immersive visualization.


\end{abstract}


\section{Introduction}
Software analytics aims to obtain insightful and actionable information about software systems, projects, and users based on the analysis and visualization of data from software artifacts~\cite{Zhang2013}.
Three-dimensional (3D) software visualizations have existed for many years, which also applies for their use in software analytics.
However, they often fall short of expectations.
This is attributed in particular to the areas of navigation, selection, occlusion, and text readability~\cite{Brath2014}.
Merino et al.~\cite{Merino2018} showed in a controlled experiment that these issues can be minimized by displaying 3D visualizations in immersive augmented reality (AR).
Navigation and occlusion in particular pose fewer problems in AR.
The authors also found that software comprehension tasks with 3D software visualizations could be solved more effective with AR devices than standard computer screens and identified great potential of AR for software engineering~\cite{Merino2020}.
Nevertheless, software analytics is hardly ever done in AR and in our own experience many software developers approach this topic with reservations.   

With this work, we contribute to the understanding of  problems in augmented software analytics. 
Based on the findings of Merino et al.~\cite{Merino2018}, we deal with the research question: 
\textit{``Which parts of the user interface of software analytics applications in augmented reality cause usability problems?''}
    
To answer this question, we developed an AR application for software analytics, which corresponds to the current state-of-the-art.
It reimplements the city metaphor, originally introduced in~\cite{Wettel2008c}, which is a one of the most popular 3D software visualizations.
We conducted a qualitative study with this application to identify usability problems in AR.
We could show that most problems are not caused by the city visualization itself, but by using inappropriate interactive user interface elements such as dialog boxes, and scrollbars.
The main contributions of the paper are the identification of usability issues in AR and the discussion how to solve them.
We can also confirm many of the findings of Merino et al.~\cite{Merino2018}, for example, that the usability problems are mainly caused by selection and not by navigation and occlusion.

\section{Related Work}
There are some approaches using AR to support the analysis and visualization of data from software repositories.
Souza et al.~\cite{Souza2012} present SkyscrapAR, one of the first tools using the city metaphor to investigate the evolution of a software system in AR.
Kapec/Brndiarov~\cite{Kapec2015} introduce a graph-based visualization of software systems on a see-through display.
Sharma et al.~\cite{Sharma2018} describe an AR tool for managers to display developer information, such as monthly developer time-line, code commit statistics, and recently owed technical debts.
A further approach of these authors~\cite{Sharma2019} outlines the concept of smart immersive software engineering workspaces where data from different sources are integrated and relevant information is presented via AR devices to developers.
Reipschl{\"a}ger et al.~\cite{Reipschlaeger2018} present DebugAR, a tool to debug software systems in AR.
It has also been studied, how to control the exploration of software architectures in AR through speech~\cite{Seipel2019a,Seipel2019b,Schreiber2019}.
Mehra et al.~\cite{Mehra2019} present XRaSE, a tool to create immersive representations of software systems to support comprehension.
Although they sketch a planned evaluation study design, we could actually find one approach that systematically investigates usability issues in AR.
Merino et al.~\cite{Merino2018} conducted a controlled experiment and a user study.
They found that immersive AR facilitates navigation, reduces occlusion, and improves user experience without affecting the performance.
But selection and text readability still remain open issues.
Here, we continue with our research.


\section{Study Design}

We use the goal definition framework~\cite{Basili1988} to summarize the scope of our experiment:
\begin{framed}
        Analyze \textit{augmented reality applications for software analytics}
        for the purpose of \textit{evaluation}
        with respect to \textit{usability}
        from the point of view of \textit{software developers}
        in the context of \textit{software developers performing comprehension tasks}.
\end{framed}

We used a Microsoft HoloLens headset with a stereo 1268~x~720 resolution, 60 Hz content refresh rate, and 30\degree H and 17.5\degree V field of view.
The application uses the Mixed Reality Toolkit (MRTK), which is based on Unity 3D.
This is the de facto standard for HoloLens applications.
To focus on the user interface, we have integrated our application into the Getaviz framework, which can create city visualizations for Java, C\#, and Ruby programs~\cite{Baum2017}.
We import the output of Getaviz into Unity 3D and thus provide an alternative user interface for Getaviz.

\begin{figure}[!bt]
  \centering
  \includegraphics[width=0.45\textwidth]{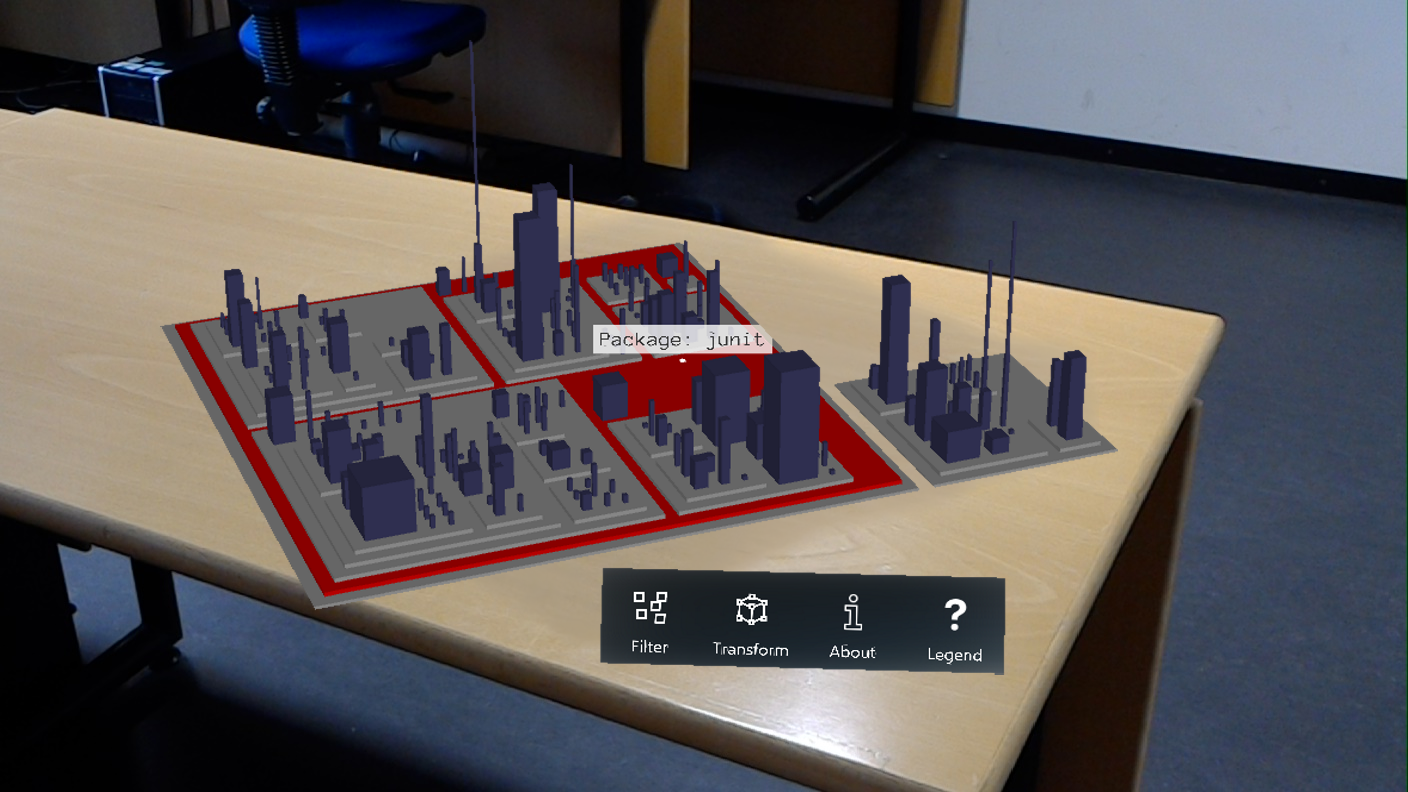}
  \caption{City visualization and app bar superimposed on a table }
\label{fig:city} 
\end{figure}

We have selected the city metaphor as visualization technique because it has proven to be effective to support software comprehension tasks on standard computer screens~\cite{Wettel2011} and in AR~\cite{Merino2018}.
The provided functionality and user interface correspond to other implementations of the city metaphor for virtual and AR.
The tool is available at the Microsoft Store\footnote{\url{https://www.microsoft.com/en-Us/p/getaviz/9mx6v3dt7p3s}}.
Additionally, its usage is demonstrated in a screen-cast\footnote{\url{https://www.youtube.com/watch?v=Egb1TBfbWDw}}.
The user interface consists of the city itself and an app bar with buttons as shown in Fig.~\ref{fig:city}.
Class names are displayed as tooltips by hovering over the building.
The source code of a class can be viewed in a separate dialog window.
Users can navigate freely, this means, they can rotate and move the visualizations or zoom in and out.
Similarly, they can navigate the system by walking and approaching buildings in the city.
To be able to study more complex interactions, we have added a filter dialog.
The user can navigate through the package containment tree and select individual packages.
In the future, districts can thus be hidden.
Currently, this window is only a mock-up and does not affect the city visualization.

\subsection{Tasks}

Each subject had to perform a number of tasks.
These were chosen to cover the complete functionality of the application.
This includes recognizing visual patterns and identifying outliers.
However, we have taken care to ensure that non-analytical aspects of use are also covered by the tasks.
This includes support tasks such as setting up the visualization, starting dialog windows, and using voice commands.
We included these tasks because they are essential for practical use and may also cause usability problems.
Table~\ref{tab:tasks} gives an overview of all tasks.
Each task belongs to one of the following categories.

\begin{itemize}
 \item \textbf{Help \& Setup:} 
    This category contains all support tasks.
    This includes starting and setting up the application and help menus.
    Problems with these tasks make further use considerably more difficult.
 \item \textbf{Visualization:}
    For these tasks the subject has to work with the visualization in its current state.
    This includes above all the reading of information and the recognition of patterns.
    Only minimal interaction with the application is necessary, since neither information has to be filtered nor additional information has to be displayed.
\item \textbf{Dialog:}
    For these tasks, the subject must change the state of the visualization by hiding information or showing additional information.
    This requires complex interactions and additional dialog windows.
\end{itemize}

\begin{table*}[bt]
\caption{Overview of all tasks}
\label{tab:tasks}
\centering
\resizebox{\textwidth}{!}{\begin{tabular}{llp{13  cm}ccc}
  \toprule
  Category & ID & Task & Evaluation I & Evaluation II & Evaluation III \\
  \midrule
    \multirow{3}{*}{Help \& Setup} & 
    T1 & Place the visualization on the table in front of you. & \checkmark & \checkmark & - \\
    & T2 & Complete the tutorial. & - & - & \checkmark \\
    & T3 & Change the size, orientation and position of the visualization until you feel you have an optimal view of the city. & \checkmark & \checkmark & \checkmark \\
    
    \midrule
    
    \multirow{5}{*}{Visualization} & 
    T4 & Find the class with the most methods. & \checkmark & \checkmark & \checkmark \\
    & T5 & Find the class with the fewest methods. & \checkmark & \checkmark & \checkmark \\
    & T6 & Find the class with the most attributes. & \checkmark & \checkmark & \checkmark \\
    & T7 & Find the class with the fewest attributes. & \checkmark & \checkmark & \checkmark \\
    & T8 & Is there a God class? If so, what is its name? & \checkmark & \checkmark & \checkmark \\
    
    \midrule
    
    \multirow{10}{*}{Dialog} &
    T9 & Open the filter menu with the voice command \textit{Filter}. & - & \checkmark & \checkmark \\
    & T10 & Filter for the package \texttt{org.junit.validator}. & \checkmark & \checkmark & \checkmark \\
    & T11 & Close the filter menu with the voice command \textit{Close}. & - & \checkmark & \checkmark \\
    & T12 & Look at the source code of any class. & \checkmark & - & - \\
    & T13 & Open the source code dialog by focusing on a class and using the voice command \textit{Select}. & - & \checkmark & \checkmark \\
    & T14 & Use the voice commands \textit{Up} and \textit{Down} to navigate in the source code display. & - & \checkmark & \checkmark \\
    & T15 & Now use the click gesture to navigate in the source code display. & - & \checkmark & \checkmark \\
    & T16 & Close the source code dialog by applying the click gesture to the button in the upper right corner. & - & \checkmark & \checkmark \\
    & T17 & Open the About dialog using the voice command \textit{About} and read the dialog. & - & - & \checkmark \\
    & T18 & Open and read the legend dialog. & - & - & \checkmark \\
    
    \bottomrule

\end{tabular}}
\end{table*}

\subsection{Procedure}

We applied an iterative usability evaluation approach~\cite{Nielsen1993} with a total of three iterations.
After each iteration, the feedback of the subjects regarding usability was implemented.
Each subject was given a short introduction about the city metaphor and software visualization in general.
To ensure that each subject had the same prerequisites, this introduction was given in written form.
After the introduction, the subjects should familiarize themselves with the HoloLens.
The \textit{Learn Gestures} app, which is installed on the HoloLens by default, was used for this purpose.
Thus the subjects were familiar with the gestures and handling of AR applications.
The subjects then worked through the tasks one after the other.
Feedback from the subjects was obtained in the form of a semi-structured interview.
After each task they were asked if they experienced any usability issues.
We recorded the answers of the subjects to the task set and to the supplementary questions about problems as well as conspicuous interactions with the application, for example, where the subjects positioned windows in the room or whether they used voice or gesture control.
Each session took place at the same location and was conducted by the same experimenter.
In every iteration we used the same visualization of JUnit 4.1.

Based on user feedback new dialogs were added in each iteration.
For example, in the first two iterations the visualization metaphor was described on a piece of paper.
For the third evaluation, a help dialog was added, which contains the same information.
New tasks were also added cover these additional dialogs.
The help dialog was evaluated in the third evaluation by the new task \textit{T18}.

\subsection{Subjects}

Five subjects participated in each evaluation.
None of them participated more than once.
Three subjects of the second evaluation are computer science students, the other 13 subjects are professional software developers.
They participated in the evaluation voluntarily and did not receive any compensation.
Two subjects have already had experience with AR, but not in the context of software analytics.

In the first evaluation, five male subjects with an average age of 30.4 ± 5.9 years participated.
They have an average of 9.0 ± 6.4 years experience in software development.
In the second evaluation, two female and three male subjects with an average age of 26.4 ± 3.00 years participated.
The professional software developers have an average of 5.0 ± 3.0 years experience in software development.
The students have experience with software development only within the context of their studies.
In the third evaluation, five male subjects with an average age of 30.6 ± 4.88 years participated.
They have an average of 8.8 ± 3.0 years of experience in software development.

\section{Results}

\subsection{Evaluation I}

\textbf{Help \& Setup:}
    All subjects could successfully complete \textit{T1} and \textit{T2}.
    However, the subjects were not aware that they could adjust the visualizations position and size at any time.
    It was noticeable that all users turned the visualization until the edges of the city were parallel to the table.
    Also the size was usually adjusted so that the city fitted completely on the table.
    Some subjects reported that they would have preferred to adjust the city to a larger size, but would have needed a larger table for this.
    
\textbf{Visualization:}
    All tasks were solved correctly.
    However, the subjects made a number of suggestions for improvement.
    The app bar was described several times as annoying because it obstructed viewing the visualization.
    All subjects had problems aiming at thin and small buildings.
    None of them enlarged the city afterwards to make targeting easier.
    The tooltips with class names were occasionally obscured by other buildings, so that they could not be read.
    This is an implementation error, since this is not the desired behavior.
    We fixed this by implementing a custom shader that always displayed the tooltip in front of any building.

\textbf{Dialog:}
    The source code dialog was often not found after opening. 
    The test subjects felt that the text field for the source code was too small.
    In addition, they had difficulties in operating the scrollbar.
    They often moved beyond the targeted point.
    As a consequence, we completely redesigned the dialog.
    We increased its size and added an opening animation so that the window is easier to find for the subjects.
    Additionally, we replaced the scrollbar with buttons for scrolling up and down, similar to the keys \textit{page up} and \textit{page down} on a keyboard.
    All subjects had problems with operating the filter function.
    Especially the use of the virtual on-screen keyboard caused problems.
    Therefore, we also turned them into buttons.
    Some subjects found the use of the HoloLens through gestures to be very strenuous.
    Hence, for the next evaluation, we implemented an additional voice control to reduce strain on the arms.
    
    
\subsection{Evaluation II}

    In the second evaluation, some tasks were added because the application could now be controlled by voice commands.
    Furthermore, the tasks for displaying the source code have been extended to achieve a higher coverage of the control elements.
    Overall, the feedback of the subjects after the second evaluation was noticeably more positive.

\textbf{Help \& Setup:}
    Some problems, which were also apparent in the first evaluation, reappeared.
    Some subjects had problems using the navigation mode intuitively.
    They could not recognize the translation by themselves.
    Other subjects did not recognize what the handles on the bounding box could be used for.
    To solve this problem, an interactive tutorial was implemented for evaluation III.
    It uses spoken text to ask the user tasks that will guide him through the functions of the app.
    When the user completes the task, the next task is set until the user has used all functions once.

\textbf{Visualization:}
    The subjects were still able to solve the technical tasks correctly.
    When reading the class and package names there was no negative feedback this time.
    The targeting of thin and small buildings was again difficult.
    One of the subjects asked for a tolerance range.
    This should ensure that an already targeted building is still highlighted, even if the user looks slightly to the side of the building.
    This is only possible when the cursor does not focus another building or district.
    Some of the subjects were disturbed again by the app bar because it blocked their view.

\textbf{Dialog:}
    With the revised filter dialog, all subjects were able to solve the corresponding task \textit{T10}.
    Users rated the handling of the source code dialog as good.
    They remarked that line numbers would be practical for orientation in the source code.
    The problem that subjects did not find the dialogs occurred again.
    When they were made aware of the animation, they said that they did not notice it the first time.
    Some said that they had to concentrate too much on the gestures.
    Since users had problems with the correct execution of gestures, an acoustic signal should also be played as further feedback when a button is successfully clicked.
    
\subsection{Evaluation III}
    For the last evaluation, the tasks were extended once again to be able to examine the newly added in-app tutorial and the new help menu.
    
\textbf{Help \& Setup:}
    The tutorial made it much easier to get started with the application.
    The tutorial tasks were successfully solved by all subjects without further questions.
    All users found the tutorial well structured and easy to understand.
    It was clearly visible how the tutorial helped the subjects to solve the follow-up tasks.
    
\textbf{Visualization:}
    The targeting of thin buildings was simplified by a tolerance range.
    If a user sees a building and then looks next to it without hitting another building or district, the first building is still highlighted.
    The legend helps to avoid some questions with regard to the metrics in the professional tasks.
    Due to the tolerance range it was easier for subjects to target buildings that are both thin and tall.
    However, they reported that targeting small, thin buildings is still difficult.

\textbf{Dialog:}
    After subjects in the second evaluation had problems finding the dialog with the open animation, the logic was revised again.
    The dialog is now opened at the height of the city instead of the height of the HoloLens.
    This means that the dialog often overlaps with the city, but there is always a part of the dialog in the user's field of view.
    The tutorial explains to the user how to move dialogs.
    In all dialog windows an indicator has been added in the upper left corner to show that the dialog can be moved.
    Users can now use this knowledge to position the dialog according to their wishes.
    Some of the subjects criticized the long delay of more than two seconds in voice control.
    This was particularly noticeable with the voice command \textit{down}.
    All subjects stated that the voice control worked, but that they would rather not use it.
    This was not related to the delay but to a general rejection of voice control.
    
\section{Discussion}
    
    It is striking that we did not have to make any changes to the city metaphor itself.
    Right from the start, all tasks of the visualization category could be solved completely.
    None of the subjects suggested changes to the visualization.
    
    Table~\ref{tab:issues} gives an overview of the identified usability issues.
    Most issues belong to the category \emph{Selection}.
    This includes problems with the selection of buildings and conventional user interface elements.
    Although we were able to mitigate these issues, they still occurred in the final evaluation.
    Occlusion was not a problem within the city.
    Tooltips and app bar caused occlusion, but these issues could be solved.
    In the first evaluation a too small text size was criticized which could easily be fixed.
    Apart from that, the subjects did not have any problems reading texts.
    At the beginning, only a few help functions were implemented in the application.
    Due to the feedback of the users, their number was increased with each iteration.
    This made the subjects feel much more confident in using the application.
    This finding is not surprising.
    However, it shows that previous studies have concentrated too much on the implementation of the metaphor and neglected the user guidance.
    In our study, the identified usability issues with respect to missing user guidance could be completely solved.
    The solution consisted of a tutorial and help functions that can be easily implemented.

    
    \begin{table}[!bt]
      \caption{categorization of identified usability issues}
      \label{tab:issues}
      \centering
      \begin{tabular}{l|ccc}
      \toprule
      Category & Evaluation I & Evaluation II & Evaluation III \\
      \midrule
      Text readability & 1 & - & - \\
      Occlusion        & 2 & 1 & - \\
      Navigation       & - & - & - \\
      Selection        & 3 & 1 & 2 \\
      User guidance    & 1 & 2 & - \\
      \bottomrule
      \end{tabular}
    \end{table}    
       
    Most usability problems were caused by conventional user interface elements, such as dialog windows, buttons, and scrollbars.
    In the first evaluation we used the standard components of MRTK without further customization to implement the app bar and dialog windows.
    This solution was found to be unsuitable by the subjects, so we replaced the standard components with our own implementations.
    This was necessary because, for example, MRTK's app bar cannot be adapted to the user's position as suggested by the subjects.
    The subjects liked our adapted implementations much more, so that no further suggestions for improvement were given.
    
    Our application uses MRTK 2017.4.3.0, which was the latest release at the time of the evaluations in 2019.
    The MRTK has now been fundamentally revised.
    However, the problems with the standard components are still there, for example, the position of the app bar still cannot be adjusted to the position of the user.        
    With better support for the WebAR interface, WebGL-based frameworks such as A-Frame could possibly be more suitable in the future.
    Existing web-based software analytics tools such as Getaviz already rely on A-Frame~\cite{Baum2017}.
    
    Our study suggests that voice control is only partially suitable for facilitating the work with dialog windows.
    The voice control was rejected by our subjects, although it worked well from a technical point of view.
    The virtual on-screen keyboard has also proved to be unsuitable.
    
\begin{figure}[!tb]
  \centering
  \includegraphics[width=0.45\textwidth]{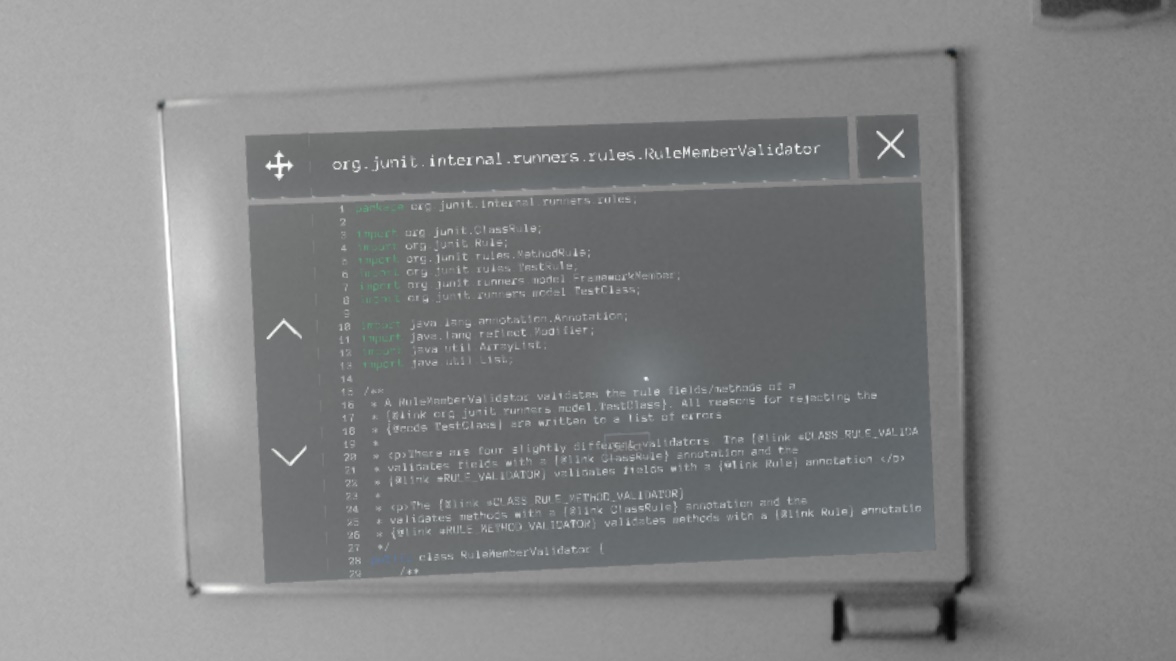}
  \caption{Source code dialog window superimposed on a physical whiteboard}
\label{fig:code} 
\end{figure}

    In previous studies the city was simply placed in space without interacting with real world objects.
    In our study, however, all subjects placed the city on a real table.
    This table was used as a reference point when the subjects moved around the room.
    They always knew where the city was and could walk around the table to view the city from all sides. 
    Some subjects also placed the source code in a way that it was displayed directly on a real whiteboard as shown in Fig.~\ref{fig:code}.
    This was not intended by us, it was a coincidence that the room in which the study was conducted had a whiteboard.
    This made it more natural for the subjects.
    We believe that this mixed reality aspect, that means, the interaction of virtual elements with real objects, is very promising.
    Greater inclusion of the real world could improve usability by requiring fewer dialog windows and buttons.
    This can be implemented within the framework of the city metaphor.
    But it is also conceivable to find a new metaphor that fits better into a typical office environment.
    
\section{Conclusion}
    
    By means of three qualitative evaluations we could show that the usability of software analytics applications for AR glasses can be improved significantly.
    There is a lot of potential for improvement especially in the area of dialog design. 
    Unfortunately, this causes a high implementation effort because available standard components are only conditionally appropriate.
    But usability can also be quickly improved through better user guidance.
    We also found indications that a stronger emphasis on mixed reality could further improve the applications.
    We will explore these possibilities in our future work.
    
\section*{Acknowledgment}

We would like to thank Stefan Mutke and the Logistics Living Lab at Leipzig University, who provided us with the used Microsoft HoloLens.

\bibliographystyle{IEEEtran}
\bibliography{literature}

\end{document}